# Experimental observation of a first-order phase transition below the superconducting transition temperature in the multilayer cuprate superconductor $HgBa_2Ca_4Cu_5O_y$

Yasumoto Tanaka[1], Akira Iyo[1], Satoshi Itoh[2], Kazuyasu Tokiwa[2], Taichiro Nishio[3], Takashi Yanagisawa[1]

[1]National Institute of Advanced Industrial Science and Technology (AIST), 1-1-1 Umezono, Tsukuba, Ibaraki 305-8568 Japan

[2]Department of Applied Electronics, Faculty of Industrial Science and Technology, Tokyo University of Science, 6-3-1 Niijuku, Katsushika-ku, Tokyo, 125-8585, Japan

[3]Department of Physics, Faculty of Science Division II, Tokyo University of Science, 1-3 Kagurazaka, Shinjuku-ku, Tokyo 162-8601, Japan

**Abstract**

A first-order phase transition is found in the multilayer cuprate superconductor, $HgBa_2Ca_4Cu_5O_y$ (Hg-1245), with a superconducting transition temperature of 108 K, under zero magnetic field. We observed a hysteretic specific heat jump around 41 K. We conclude that the Bardeen–Cooper–Schrieffer pairs have a residual entropy due to fluctuations in the phase difference between the five $CuO_2$ planes in a unit cell of Hg-1245, and that this fluctuation freezes below the first-order phase transition temperature.

## 1. Introduction

Changing one order parameter into another has been considered a special event in Bardeen–Cooper–Schrieffer (BCS) pair condensates.[1-4] It is observed in multi-component condensates, examples of which include triplet-pair condensates such as heavy-fermion superconductors[2,5-27] and superfluid helium-3.[1,7,28-38]

The conversion of the superconducting order parameter in multi-component condensates has been mainly investigated by specific heat studies, in which double specific heat jumps were observed in the absence of a magnetic field.[2,9,17,22-27] Heavy fermion triplet superconductors like $UPt_3$ are an example of multi-component condensates, with second-order phase transitions.[2,9,18] To date, first-order phase transitions in multi-component BCS pair condensates without an external gauge field, such as an electromagnetic field, or container rotation, have only been reported for superfluid helium-3.[1,32,34,38] For superfluid helium-3, the realized multiple order parameters are well understood.[29-32,34-38] Helium-3 exhibits a change of a nodal gap type order parameter with two point-nodes (Anderson–Brinkman–Morel state)[36,37] into a node-less gap type (Balian–Werthamer state).[35] Symmetrically, the nodal gap parameter does not affect the node-less parameter; because the two gap parameters do not mix or coexist. This phase transition is first-order, and there is no similar first-order phase transition in triplet superconductors.

The multiple components seen in triplet p-wave pairing originate from the degrees of freedom of spin and orbital angular momenta of the pair function. Conventionally, there is no way to realize multiple components that does not rely on the degrees of freedom of these angular momenta. Recently, however, it has been found that a multi-band and multi-layer structure can generate a new internal degree of freedom in a superconducting condensate.[39-51] In this case, the superconductivity is designated "multi-component superconductivity based on multi-band superconductors".

Multi-layer cuprate superconductors are a strong candidate for exhibiting such multicomponent superconductivity.[52-64] In Fig. 1, we schematically show a typical multi-layer cuprate superconductor, $HgBa_2Ca_4Cu_5O_y$ (Hg-1245).[65-68] This cuprate has five crystallographically inequivalent $CuO_2$ planes in its unit cell. These $CuO_2$ planes are sequentially stacked, sandwiching a Ca layer. This superconducting block is sandwiched by two $BaO-HgO_x-BaO$ charge reservoirs. The hole density in the $CuO_2$ plane neighbouring the charge reservoir layer (outer plane) is about 0.2 per Cu, whereas that in the other $CuO_2$ plane (inner plane) is about 0.06.[69,70] The superconducting transition temperature, $T_c$, of Hg-1245 is 108 K. From a muon spin resonance (μSR) study, the transition temperature for the three-dimensional antiferromagnetic order, $T_N$, is estimated to be 45 K.[71,72]

Hg-1245 exhibits several electronic bands, due to the multiple $CuO_2$ planes in one unit cell. When each band has its own superconducting quantum phase, we can consider that there are new degrees of freedom arising from the inter-band phase difference. We can simply consider that the phase differences between the $CuO_2$ planes become the new degrees of freedom.[73] The electromagnetic field couples to one of these degrees of freedom and the other internal degrees of freedom remain as electromagnetically inactive modes. There might be several nearly degenerated order parameters with an energy difference of the same scale as the interlayer pair hopping energy. A change of these order parameters is plausible. When this kind of change occurs below $T_c$, it can be observed as a specific heat anomaly.

In this article, we report a first-order phase transition below $T_c$ in Hg-1245 and discuss the dynamics of the superconducting order parameter in view of the multi-component superconductivity based on multi-band superconductors.

## 2. Experiment

A polycrystalline Hg-1245 sample was synthesized using a high-pressure technique that has been reported elsewhere.[65,66] The quality of the sample and the bulk superconductivity had been well characterized and established in our previous works.[56,67,74-79]

The sample was shaped into a plate of weight 19.6 mg. The heat capacity was measured by the thermal relaxation method using a physical properties measurement system installed by Quantum Design.[60] First, the heat capacity was measured as the temperature was reduced from 130 K to 6 K and then again while the temperature was increased. We also measured the heat capacity under an applied magnetic field of 14 T. For each measurement, the temperature of the sample stage was increased by about 2% of the measurement temperature, and then decreased. The starting temperature was lowest for each run. Before starting the run, we waited until any temperature drift had ceased. We measured the specific heat capacity three times at each measurement temperature. Except near 41 K, there are no apparent deviations in the measured data. We considered that the specific heat capacity was measured properly without any influence of latent heat around 41 K.

## 3. Results

Fig. 2 shows the overall temperature dependence of the heat capacity (C) divided by temperature (T). A hysteretic jump around 41 K is seen as well as a small jump around 108 K due to the superconducting transition at $T_c$. Although the 108 K jump (Fig. 3) does not show any apparent thermal hysteresis, the 41 K jump exhibits remarkable hysteresis (Fig. 4). The broken line corresponds to a polynomial fit to the data. In fitting, we sectioned the measurement region into several temperature sub-ranges. In Fig. 5, we plotted the difference between C/T as a function of increasing and decreasing temperature, using the fitted curve. We similarly analysed the data obtained under a magnetic field of 14 T and present the results in the same figure. The magnitude of the difference in C/T were reduced by about 30% under the applied field compared with those with no applied field. The integrated C/T is also plotted with the starting temperature of the integration set at 45 K. This gives some indication of the entropy difference between the higher and lower temperature phases. As the temperature decreases, the system gradually switches to a different phase, taking a much longer time than the relaxation time for the heat capacity measurement. Not obtained from these experiments is the variation in the transition entropy accompanied by the latent heat at 41 K with increasing temperature. The given entropy difference is then a lower limit of the actual entropy difference between the higher and lower temperature phases.

## 4. Discussion

In this section, we argue that the hysteresis of the specific heat originates in the superconducting state. We then compare the observed first-order phase transition with other phase transitions below $T_c$ found in conventional magnetic superconductors,[80-90] triplet superconductors[2,9-27] and superfluid $^3$He.[1,28,32,34-38] The key issue is the multiple components of superconducting condensates. We propose that we can explain the first-order phase transition of Hg-1245 in view of the multi-component superconductivity based on multi-band superconductors. We also briefly discuss the relationship to other phase transitions in other (non-superconducting) systems.

There is the thermal hysteresis in specific heat between 20 and 41 K. The sample does not show any minor phases causing such a thermal hysteresis by X-ray diffraction measurement.[72,91,92] The thermal hysteresis means there are two different states having different free energies, which is a typical situation giving rise to the first order phase transition. The state with increasing temperature is in a superheating state (or that with decreasing temperature is in a supercooling state) and there are difference in entropy between two states. The state with increasing temperature can be considered connecting the lower temperature state and the state with decreasing temperature can be considered connecting the higher temperature state.  First, we quantitatively analyze the entropy relevant to this phase transition. The lower limit for the entropy difference between the higher and lower temperature phases is 0.034 $k_B$ (where $k_B$ is the Boltzmann constant) per unit cell, which is too small to attribute to the magnetic entropy. For 0.6 holes per unit cell,[69] the entropy difference per BSC pair is then about 0.028 $k_B$. We argue that if the difference in entropy comes from the superconductivity, then that difference can be suppressed by a magnetic field, and indeed, a suppression is seen in Fig. 5 where a magnetic field of 14 T reduces the difference in heat capacity by about 30%. In addition, the normal region is at least 7–12% of the volume of the overall system, if we assume an upper critical magnetic field for Hg-1245 of 120–200 T[93-95] as a typical value for an ordinary cuprate superconductor having a $T_c$ of about 100 K. Thus, plausibly, the suppression could be due to a reduction in the superconducting order parameter.

In a conventional superconductor, the order parameter consists of a single component. The order parameter cannot change to a different one below $T_c$ and there is no further transition below $T_c$ in the absence of a magnetic field. For some conventional magnetic superconductors, another phase transition temperature below $T_c$ is sometimes observed.[81,82,86-90,96-99] At this phase transition temperature, magnetic ordering emerges. The emerged magnetic ordering weakens or kills the superconducting order parameter, because even in conventional magnetic superconductors the order parameter is still composed of a single component for the superconducting part. Until the superconducting order parameter is suppressed completely, the magnetic phase transition is of second order. The suppression in the amplitude of the superconducting order is observed as a change in an upper critical field $H_{c2}$, as reported for $Tb_{1.2}Mo_6S_8$.[97,99,101,102] This is a second-order phase transition and it does not destroy superconductivity. The magnetic transition becomes first order in $ErRh_4B_4$[87,107] and $Tm_2Fe_3Si_5$,[88,104,105] but the superconductivity is completely destroyed in this case. Below the first-order phase transition, Hg-1245 is in a superconducting state, in contrast with these magnetic superconductors. Its first-order phase transition cannot be considered to be consistent with a conventional magnetic superconductor.

According to Landau's free energy formula, which is a polynomial expansion in terms of the order parameters, the mixing of the order parameters for a system having two order parameters leads to a second-order phase transition.[106] If the two order parameters do not co-exist, the dominance of one suppresses the other entirely, which produces a first order phase transition. The second-order phase transition observed in magnetic superconductors corresponds to the mixing of the magnetic order and the superconductivity. The destruction of the superconductivity means the superconductivity is replaced with the magnetic order when the superconductivity has only one component.

If there is a multi-component superconducting order parameter, a change in the superconducting order parameter, rather than its destruction through the interplay of magnetism and superconductivity, becomes plausible.[6,9-13,107] For example, in the heavy fermion superconductor UPt$_3$, the Bardeen–Cooper–Schrieffer (BCS) pairs have an internal degree of freedom giving rise to multiple order parameters.[9-13] The superconductivity selects one of these order parameters, preferring antiferromagnetic order at T$_c$. The other superconducting order parameter, which as yet has not been specified definitively,[16,108-110] is selected at a lower temperature.[14,15,17] However, even in this case and for other triplet superconductors such as U$_{0.9784}$Th$_{0.0216}$Ge$_{13}$, the phase transition has still been found to be second-order experimentally,[2,9,18,27] though a first-order phase transition in triplet superconductors is predicted theoretically.[10-13]

A first-order phase transition was experimentally observed in superfluid $^3$He and the conversion of the BCS pair function established.[1,28,29,31,34-38] This is a more likely analogy for Hg-1245. Hg-1245 can be considered a type of singlet d-wave superconductor similar to other cuprate superconductors; however, we need to investigate the origin of the internal degrees of freedom that the conventional cuprate superconductors do not have. Instead of the internal degrees of freedom seen in superfluid $^3$He, Hg-1245 has internal degrees of freedom stemming from the multiple CuO$_2$ planes in one unit cell.[52,56,65,67,69,73-75] The phase difference between the five CuO$_2$ layers can be considered a new internal degree of freedom.[73] The antiferromagnetic order, established around 45 K, couples to this internal degree of freedom. The multilayer structure, the size of which is much smaller than the magnetic penetration depth, enables multicomponent superconductivity to arise, much like for a pseudo-multiband superconductor,[46,47,61-63,111-113] where a layer corresponds to a band in the multi-band superconductor. This was experimentally demonstrated in an artificial aluminium bi-layer structure by Moler's group in 2006.[49] Hg-1245 reproduces this situation using the different multiple $CuO_2$ planes in one unit cell.[111-113] Loss in the phase coherence between layers is known in a conventional cuprate superconductor in terms of the crossover from three dimensional behavior to two-dimension behavior with presence of the vortices.[114,115] In this situation, the layers are separated by the charge reservoir layer. In Hg-1245 and super-multilayer cuprate superconductor having more than 5 $CuO_2$ planes in one unit cell, the similar loss in the phase coherence between layers separated by a Ca layer and/or other $CuO_2$ layers was reported.[111] The interlayer phase fluctuation would be plausible even without vortices. We can also consider that the multiband structure can be formed by the multilayer structure, as seen in band calculations.[52]

The internal degree of freedom due to the inter-band phase difference in multi-band superconductors was first addressed by Kondo.[116,117] He showed that in a two-band superconductor, a sign reversal configuration between the two bands is a possible order parameter, aside from the same sign configuration for both bands that had generally been considered. The sign reversal configuration corresponds to an inter-band phase difference of π. The same sign configuration corresponds to zero phase difference. Later, Leggett identified the importance of the inter-band phase fluctuation in a two-band superconductor.[30,39] We can apply this concept to multilayer cuprate superconductors.

At higher temperatures, fluctuations in the inter-band/inter-layer phase difference gives a finite entropy, S, which decreases the free energy through the entropy term, -TS. Experimentally, the jump in $C/T$ at $T_c$ is about 0.01–0.02 J $K^{-2}$ $mol^{-1}$, and is less than half that for optimally doped $YB_2Cu_3O_{6.92}$ (0.035–0.06) and less than one third that for $YB_2Cu_3O_{7.0}$ (0.06–0.07).[118-123] The smaller size of the jump suggests that some portion of the degree of freedom for the BCS pairs does not contribute to the jump. The number of holes can be estimated to be about 0.45 including the CuO chain contribution for $YB_2Cu_3O_{7.0}$ and about 0.6 for Hg-1245. The reduction in entropy due to the pre-formed pairing, as described for $YB_2C_3O_{7.0-\delta}$,[118–121] cannot fully account for the source of the observed suppression of the specific heat jump. The residual part from the degree of freedom of each pair not contributing to the jump at $T_c$ contributes to the interlayer phase fluctuation inside the unit cell. Energy scale of the interlayer phase fluctuation inside the unit cell would be comparable to or smaller than interlayer fluctuation sandwiching the charge reservoir layer.[77,78,111] The latter fluctuation is seen as a broad future of the specific heat jump in the conventional high-anisotropic cuprate superconductor such as $Bi_{2.12}Sr_{1.9}Ca_{1.08}Cu_{1.96}O_{8+x}$ (Bi-2212).[123] This two-dimensional behavior can be considered the loss in the interlayer coherence. Hg-1245 shows a similar broad future of the specific heat jump because Hg-1245 has a high anisotropy.[77,78,111,124] The interlayer phase fluctuation in the unit cell can be also large. For an ordinary phase fluctuation directly couples with the thermal bath through the electromagnetic field.[125,126] In addition, there is a gap for a collective excitation by the coupling with the electromagnetic field.[127-129] In contrast to this inter-unit cell phase fluctuation, the phase fluctuation inside the unit cell and the inter-band phase fluctuation tend to be free from the electromagnetic field in a simply connected specimen, which means there is no holes.[41,42,130] Its gap for a collective excitation which is Leggett mode, is governed by the interlayer Josephson coupling inside the unit cell. It plausibly survives down to low $T_c$ without large dissipation.

The interlayer phase fluctuation inside the unit cell reduces the interlayer Josephson coupling energy. As that temperature decreases, the entropy term weakens and the Josephson coupling becomes dominant. The inter-layer phase difference might then be locked. Antiferromagnetic order might favour this locked-phase superconductor.[131] A mathematically equivalent situation to phase-difference locking was discussed for three-component frustrated superconductivity, which is an extension of the Kondo–Leggett schema.[41,132-146] This theory

predicts that the phase-locking transition is first-order and has remarkable hysteresis.[146]

Thermal fluctuations associated with the internal degrees of freedom can be expected for conventional triplet superconductivity and $^3$He superfluidity. However, the temperature is too low for such fluctuations to emerge. States having such thermal fluctuations need not be considered, though collective excitation[28,147-155] by an external field and topological excitation have been actively studied.[155-162] Hg-1245 has a high superconducting transition temperature and a high-temperature first-order phase transition, both much higher than that of a triplet superconductor and $^3$He superfluidity. The interlayer phase difference fluctuation must also be considered.

The fluctuation of the phase and phase-locking transition of the order-parameter has been considered in the Peierls transition observed in tetrathiafulvalene(TTF)-7,7,8,8-tetracyanoquinodimethane(TCNQ) and tetraselenafulvalene(TSF)-TCNQ organic conductors.[163-169] In this transition a charge density wave (CDW) becomes the order parameter. There is a successive growth of coherence of the CDW from one-dimensional to three-dimensional order.[170] At higher temperatures, the CDW develops in a one-dimensional column, in which TTF, TSF or TCNQ molecules stack. As the temperature decreases, three-dimensional order is established by an inter-columnar interaction. The relative locations of CDWs on the different columns are locked and CDW motion freezes. The inter-layer Josephson interaction (or inter-band pair interaction) in Hg-1245 corresponds to the inter-columnar interaction in TTF-TCNQ. Ordinary pair interactions (or intra-band pair interactions) in the $CuO_2$ planes correspond to the intra-columnar interactions (electron–phonon interactions) in the Peierls system. However, the successive growth of the coherence in an ideal multi-band superconductor is not relevant to the spatial dimensionality, unlike in a real Peierls system, and instead is a "dimensional crossover" that occurs in the internal space created by the band structure and the strong contrast in the strength of the pair interaction, in other words, the interband interaction is much weaker than the intraband interaction.

There is considerable variety in the phase transitions in multi-order parameter systems such as antiferro–antiferro, antiferro–ferro magnetic cooperative and competing systems.[171-177] Even in some insulators, competition and cooperation between different symmetrical crystal distortions results in multiple transitions.[178] The new transition found in Hg-1245 can be accepted as one such situation. Once superconductivity has an internal space and multiple components, fluctuation inside the internal space alters the schema of the superconductivity. Conventional BCS theory was based on the assumption that a BCS pair does not have any entropy,[179-180] in other words, there is no fluctuation inside the internal space of the pair. Of course, a conventional superconductor has the superconducting quantum phase as an internal phase. But, fluctuations inside this internal space are killed by the coupling to the electromagnetic field known as spontaneous symmetry breaking.[127-129] The first-order phase transition found in Hg-1245 suggests that neither the basic assumption of BCS theory or spontaneous symmetry breaking need always be fulfilled in a general multi-band superconductor. This finding expands the range of possible superconductors awaiting discovery.

We note also that this finding does not exclude the possible emergence of other new superconducting phases which may also explain the first order transition observed in this study.[181] The possibility of odd frequency superconductivity, in which the sign of the gap parameter changes across a Fermi wave vector under the staggered field has been noted.

## 5. Conclusion

We observed a first-order phase transition below $T_c$ in Hg-1245, which might be a d-wave singlet superconductor. The phase difference between the three crystallographically inequivalent kinds of $CuO_2$ plane becomes an internal degree of freedom. The first-order phase transition suggests a new type of change in the order parameter involving this internal degree of freedom. If so, this superconductor would be a new, designable solid-state example of superconductivity having a non-Abelian gauge field based on a multiple component quantum phase, which is also a key topic in particle and cosmological field theory.[45,182-190]

Figure Captions

Fig. 1. (Color online) Schematic crystal structure of multi-layer cuprate superconductor, Hg-1245 generated by a three-dimensional visualization program for structural models, "Visualization for electronic and structural analysis" (VESTA 3) [68].

Fig. 2. (Color online) Overall behaviour of C/T (specific heat (C) divided by temperature (T)) for Hg-1245. The open circles mark C/T data obtained while decreasing the temperature and the solid circles mark C/T data obtained while increasing the temperature. The scale of C/T without an external magnetic field is marked on the left axis. That with a magnetic field of 14 T is marked on the right axis. The vertical arrows indicate two anomalies. The higher temperature anomaly at 108 K is due to the superconducting transition. The lower anomaly shows a large hysteresis around 41 K.

Fig. 3. (Color online) Jump in the C/T curve at $T_c$, exhibiting no hysteresis. The dotted lines are interpolations obtained by polynomial fitting. The symbols and vertical axes are as in Fig. 2.

Fig. 4. (Color online) Anomaly in the C/T curve at the lower transition temperature at 41 K. The large thermal hysteresis observed is a first-order phase transition. The symbols, lines and vertical axes are as in Figs. 2 and 3.

Fig. 5. (Color online) Thermal hysteretic difference in C/T curves without a magnetic field (solid line) and under an applied magnetic field of 14 T (dotted line). The scale is marked on the right. Both sets of data were deduced from interpolated data. The magnetic field suppression of thermal hysteresis supports the hypothesis that the anomaly originates in the superconducting part of the order parameter, because, under an applied magnetic field, the superconducting area decreases through the formation of normal cores. The integrated C/T without a magnetic field is shown (the corresponding scale is on the left). The starting temperature of the integration is set at 45 K, which is just above the first-order transition. This gives only a lower limit for the difference in the entropy, because there is an additional transition entropy accompanied with the latent heat at the phase transition that our relaxation technique cannot measure.

Figures

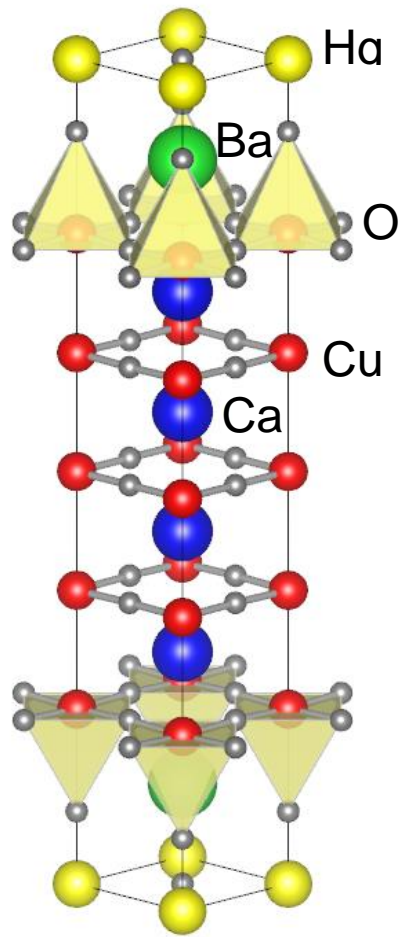

Fig. 1.

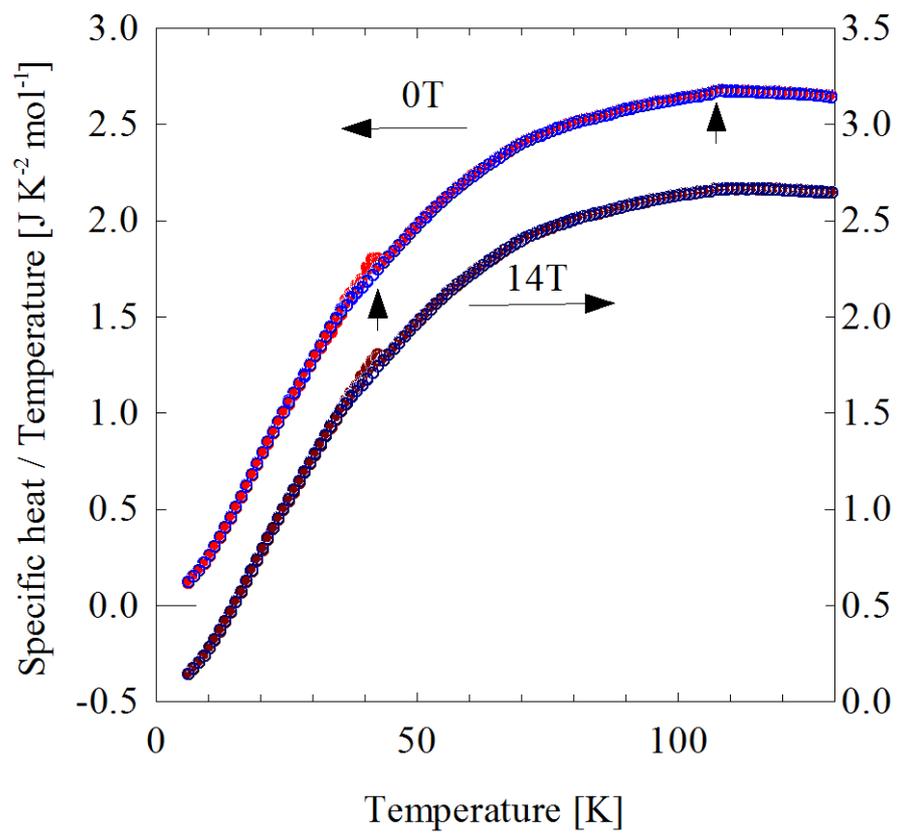

Fig. 2.

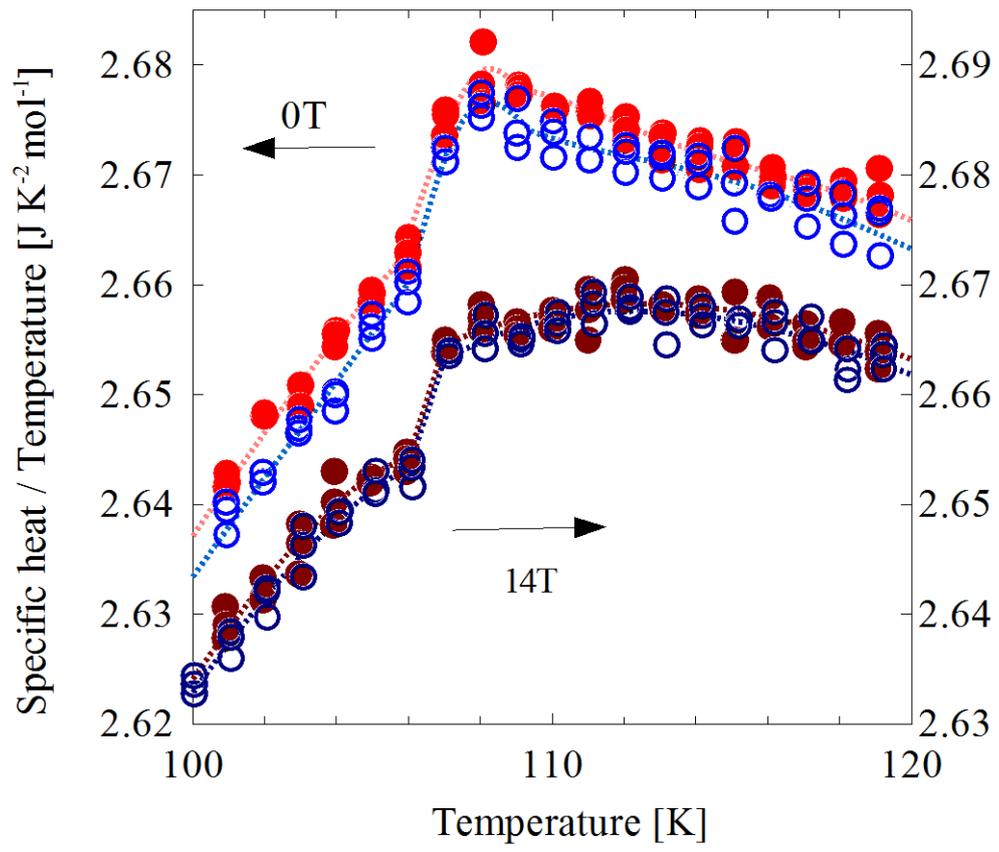

Fig. 3.

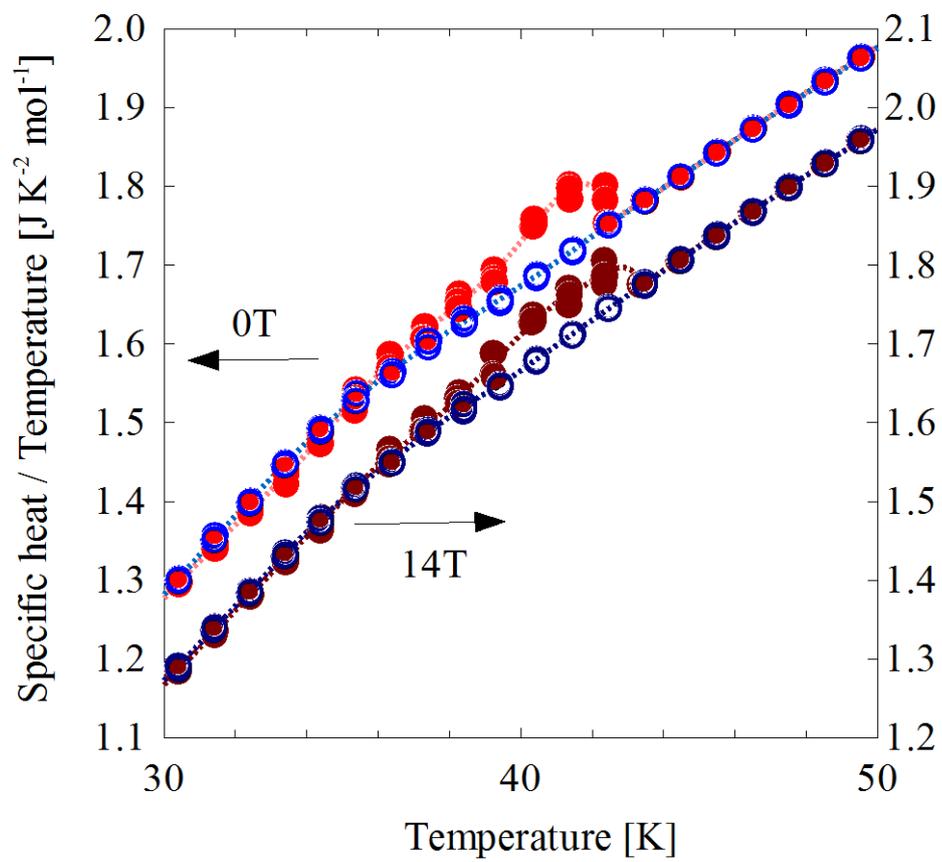

Fig. 4

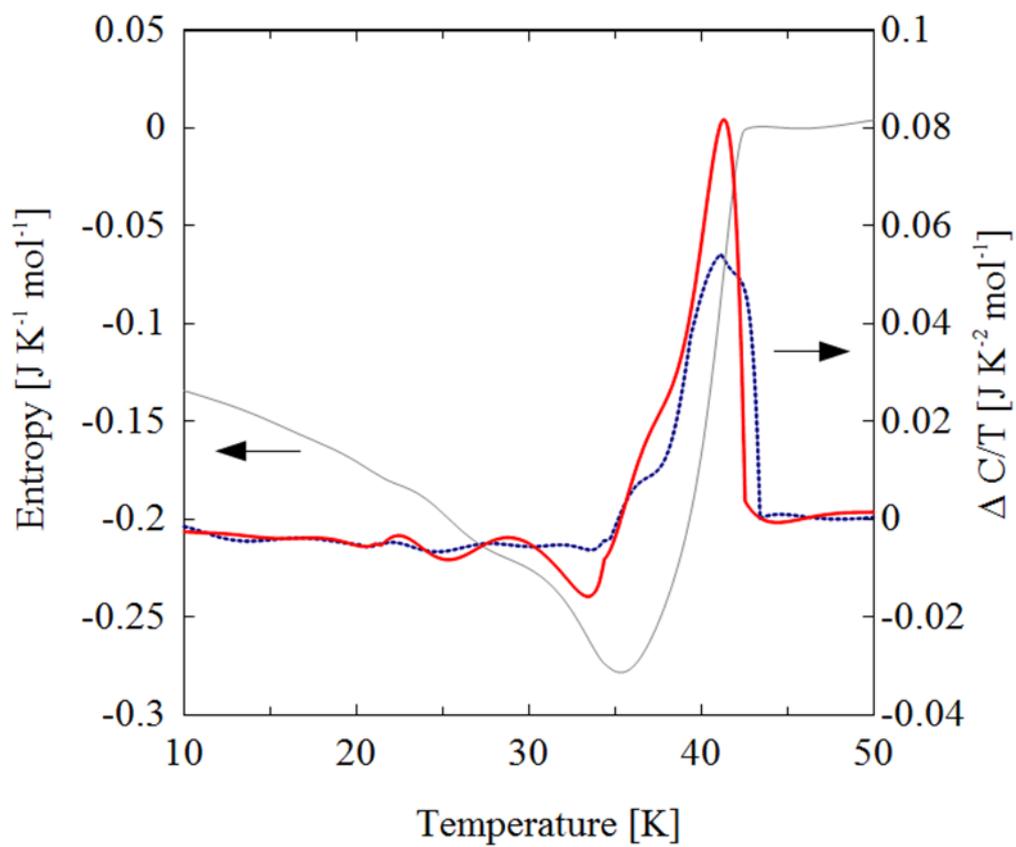

Fig. 5.